\documentclass[aip,reprint]{revtex4-1}
\usepackage{graphicx,color,amsmath}

\begin{document}

\title{Types of quantum turbulence} 

\author{C. F. Barenghi}
\email{carlo.barenghi@newcastle.ac.uk}
\affiliation{Joint Quantum Centre (JQC) Durham-Newcastle, School of Mathematics, Statistics and Physics, 
Newcastle University, Newcastle upon Tyne NE1 7RU, United Kingdom}

\author{H. A. J. Middleton-Spencer}
\email{h.a.j.middleton-spencer2@ncl.ac.uk}
\affiliation{Joint Quantum Centre (JQC) Durham-Newcastle, School of Mathematics, Statistics and Physics, 
Newcastle University, Newcastle upon Tyne NE1 7RU, United Kingdom}

\author{L. Galantucci}
\email{l.galantucci@iac.cnr.it}
\affiliation{Istituto per le Applicazioni del Calcolo M. Picone, IAC-CNR, 
Via dei Taurini 19, 00185 Roma, Italy}

\author{N. G.Parker}
\email{n.g.parker@newcastle.ac.uk}
\affiliation{Joint Quantum Centre (JQC) Durham-Newcastle, School of Mathematics, Statistics and Physics, 
Newcastle University, Newcastle upon Tyne NE1 7RU, United Kingdom}

\date{\today}

\begin{abstract}
We collect and describe the observed geometrical and dynamical properties of 
turbulence in quantum fluids, particularly superfluid helium and
atomic condensates for which more information about turbulence is available.
Considering the spectral features,
the temporal decay,
and the comparison 
with relevant turbulent classical flows, we identify
three main limiting types of quantum turbulence:
Kolmogorov quantum turbulence, 
Vinen quantum turbulence, and strong quantum turbulence.
This classification will be useful to analyse and interpret new 
results in these and other quantum fluids. 
\end{abstract}

\pacs{}

\maketitle 


\section{Introduction}
\label{sec:intro}

Turbulence, ubiquitous in nature and technology, is still a major 
problem of classical physics. 
As new contexts of this problem are investigated,
turbulence is better understood and new physics is discovered. 
A context of current interest is turbulence in quantum fluids, or
{\it quantum turbulence}, the study of which was pioneered by the late
W.F. Vinen \cite{Vinen-counterflow}.   Quantum fluids are fluids that, due to
Bose-Einstein condensation of the consistuent particles, exhibit 
quantum mechanical effects at the macroscopic level.
Example systems are liquid helium (bosonic $^4$He and fermionic $^3$He), 
ultracold atomic gases, polariton condensates, magnons, electrons in metals,
the interior of neutron stars and models of dark matter. 

\begin{figure}[!h]
{\includegraphics[width=0.4\textwidth]{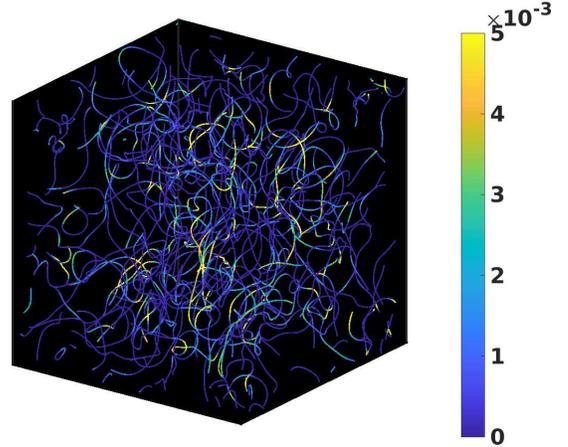}}
\caption{\label{fig:1} 
Snapshot of quantum turbulence in superfluid helium
(corresponding to the Vinen regime, see
Section~\ref{sec:Vinen}) at $T=1.9~\rm K$
computed using the VFM in a cube of size $D=1~\rm cm$. Because of the periodic
boundary conditions used, vortex lines which seem to terminate at a
boundary continue on the opposite side of the cube. Note that this image
does not display the orientation of the vortex lines, and that the lines'
thickess is exaggerated for the sake of clarity (in this numerical simulation,
the average inter-vortex distance, $\ell=9.7 \times 10^{-2}~\rm cm$, 
is many orders of magnitude larger than the helium vortex core radius, 
$a_0 \approx 10^{-8}~\rm cm$.)
The vortex lines are colour-coded according to $\vert h(\zeta) \vert $,
where $h(\zeta)$ is the helicity density.
}
\end{figure}

The main property of quantum fluids \cite{Primer}
is that their vorticity is confined to
individual ({\it discrete}) vortex lines of fixed circulation 
$\kappa=h/m$, where $h$ is Planck's constant and $m$ the mass
of the relevant boson. This quantization of the vorticity, conjectured 
by Onsager \cite{Onsager} and
experimentally demonstrated in liquid helium by Vinen \cite{Vinen-quantum}, is a fundamental property
which arises from the existence
of a governing macroscopic wavefunction. It also makes quantum turbulence
visibly different from ordinary (classical) turbulence:
whereas classical turbulence
contains a {\it continuous} distribution of eddies of arbitrary sizes 
and strengths, quantum turbulence consists of a disordered tangle of
{\it discrete} vortex lines of fixed circulation and thickness, as shown in
Figure~\ref{fig:1}.


A second property of quantum fluids is the two-fluid nature \cite{Primer}.    
The Bose-Einstein condensate, which exists when the temperature of 
the system is below a critical temperature,  constitutes a fluid which is free from viscosity.
At non-zero
temperature, this condensate coexists with an incoherent thermal part,
which plays an important role in turbulence as it dissipates
the kinetic energy of the vortex lines. In liquid helium at sufficiently
high temperature the mean free path of thermal excitations is short
enough to form what is effectively a classical fluid called
the normal fluid component.  Liquid helium therefore becomes a mixture
of an inviscid superfluid (associated to the condensate) 
and a viscous normal fluid as described by Landau's
two-fluid theory. The thermal excitations (phonons
and rotons) are scattered by the velocity field of the vortex lines, 
creating a mutual friction force between normal fluid and superfluid 
components.  When stirred, both fluid components
may become turbulent, creating an unusual doubly-turbulent state 
\cite{Melotte,Guo-excimers}
consisting of continuous normal fluid eddies and discrete superfluid vortex lines
which interact with each other.

Despite these differences between quantum fluids and ordinary fluids,
experiments with turbulent liquid helium have revealed that, under certain conditions, there are remarkable
similarities between quantum turbulence and classical turbulence.
But under other conditions, quantum turbulence is wholly unlike
classical turbulence.  

As the study of quantum turbulence is currently being
extended from superfluid helium to atomic Bose-Einstein condensates and
other quantum fluids, the existence (or not) of a classical limit 
for these out-of-equilibrium quantum systems is an important problem.
At this stage, it is therefore useful to critically review the main observed
properties of turbulence in quantum fluids, to consider the conditions 
under which these properties appear, and to 
compare the different regimes of quantum turbulence with each other and with classical turbulence. Putting together
information which is scattered over the literature, our aim in this paper
is to identify and characterise the types of quantum turbulence which currently exist.    

 
\section{Phenomenology of quantum turbulence}
\label{sec:phenomenology}

Experimental, theoretical and numerical studies suggest that the 
phenomenology of quantum turbulence can be organized in the
following main types:
(i) {\it Kolmogorov quantum turbulence}, (ii) {\it Vinen quantum turbulence},
and (iii) {\it strong quantum
turbulence}. These names and this classification are motivated
by comparisons with classical turbulence. We stress that we limit our
attention to the quantum fluids which, until now, 
have received most of the 
attention from the point of view of turbulence: 
superfluid $^4$He, superfluid $^3$He-B, and atomic condensates.
Before we describe these three types
of quantum turbulence, we recall the basic properties of classical
turbulence, which serves as a reference.


\subsection{Classical turbulence}
\label{sec:classical}

The paradigm of classical turbulence is the
statistically-steady, homogeneous and isotropic turbulence (HIT) of a viscous 
incompressible fluid \cite{Frisch}. 
This state of disorder is achieved, by continual supply of kinetic energy and removal of heat
to compensate for the dissipation arising from the fluid's viscosity,  for example,
in a wind tunnel.
Theoretically and numerically, classical turbulence is studied by 
solving the Navier-Stokes equation. HIT is the conceptual
pillar which supports the engineering toolkit for engineering
applications (the $k-\epsilon$ model and Large Eddies Simulations);
the theory of HIT makes successful predictions even outside the range 
for which it was originally designed \cite{AlvesPortela}, 
e.g. inhomogeneous and unsteady flows.  The properties of HIT are 
therefore the natural reference when considering the
properties of turbulent quantum fluids.

Classical HIT has two characteristic length scales: the
large length scale $D$ at which kinetic energy is injected
and the small length scales $\eta$ (called the Kolmogorov length scale) 
at which kinetic energy is dissipated by viscous forces. 
Vortices (eddies), continually created at large length scales, 
are unstable and break up into smaller and smaller
eddies to which they transfer their kinetic energy. It is convenient
to introduce the wavenumber $k=2 \pi/r$ corresponding to eddies of
size $r$.  In the inertial range of wavenumbers 
$k_D=2 \pi/D \ll k \ll k_{\eta}=2 \pi/\eta$, the energy transfer
is self-similar, viscosity playing no role.  Here the turbulence displays the celebrated kinetic energy
spectrum $\hat{E}(k) \sim k^{-5/3}$ (called the Kolmogorov spectrum)
which is interpreted as the signature of
a dissipationless energy cascade from large eddies to small eddies. For $k \gg k_{\eta}$
viscous forces dominate the dynamics and
the energy spectrum decays exponentially with $k$. 
Whereas the kinetic energy
is concentrated at the large length scale $D$, the vorticity is concentrated
at the small length scale $\eta$.  In the inertial range of wavenumbers,
the distribution of enstrophy $\hat{\Omega}(k)$ (vorticity squared) scales as
$\sim k^{1/3}$, peaking at $k_{\eta}$ and then decaying for 
$k> k_{\eta}$.  

The intensity of
the turbulence is quantified by the non-dimensional Reynolds 
number ${\rm Re}=UD/\nu$, where $U$ is the flow speed at the large
length scale $D$ and $\nu$ is the fluid's kinematic viscosity. 
The Reynolds number estimates the ratio of inertial and viscous forces
in the Navier-Stokes equation.   It can also be used to express the extent
of the linear separation between the smallest length scale and the
largest length scale according to $\eta/D \approx {\rm Re}^{-3/4}$. 

Another property of HIT that is relevant to quantum turbulence
is that the distributions of the
values of velocity components are Gaussian, as confirmed by experiments
\cite{Noullez} and numerical simulations \cite{Vincent} 
(velocity increments however follow power-law statistics at small scales). 
Finally, in
classical turbulence the rate of the kinetic energy dissipation, $\epsilon$,
tends to a non-zero constant in the limit $\nu \to 0$ (i.e. 
the limit ${\rm Re} \to \infty$). This result is in sharp
contrast to what happens in laminar flows, where the dissipation
goes to zero with the viscosity. Physically, in a turbulent flow
the reduction of the viscosity is compensated by the creation of motions 
at smaller length scales containing much vorticity but little
energy. This property (called the dissipation anomaly \cite{Sreeni}) is clearly
relevant to the task of comparing classical and quantum turbulence
because superfluids have zero viscosity.

We note that, although the scenario which we have described is robust,
it is only a first approximation: higher
order statistics depart from the Kolmogorov scaling, representing 
intermittency corrections \cite{Frisch} which are beyond the scope of this paper.


\subsection{Kolmogorov quantum turbulence}
\label{sec:Kolmogorov}

There is general consensus\cite{BarenghiLvovRoche} that, 
under certain conditions, quantum
turbulence takes a form which is similar to classical turbulence, and therefore
we examine this type first. 

In any turbulent quantum
fluid there is necessarily a third characteristic length scale (besides
the large length scale of the energy injection and the small length 
scale of the energy dissipation): the average distance between
vortex lines, $\ell$. In the experiments, this parameter is
usually estimated as $\ell \approx L^{-1/2}$, where $L$ is the total
vortex line density (vortex length per unit volume).  
The vortex line density can be experimentally measured in liquid helium
by a number of techniques such as second sound attenuation, 
ion trapping and Andreev reflection\cite{Skrbek2011}.
The vortex line density is often 
considered a measure of the intensity of quantum
turbulence, although, for the sake of comparison between experiments,
a better measure \cite{tough-1982} would be the dimensionless parameter
$D/\ell $.

Direct evidence of the classical $k^{-5/3}$
Kolmogorov scaling in quantum turbulence
was provided by experiments in which $^4$He was stirred by 
rotating propellers \cite{Tabeling,Salort2021} and towed grids 
\cite{Smith}, or was driven along wind tunnels \cite{Salort2010}.
The local velocity fluctuations were measured
by miniature Pitot tubes \cite{Tabeling} or cantilever anemometers
\cite{Salort2021}. This type of quantum turbulence has been called 
``quasi-classical" or {\it Kolmogorov quantum turbulence}. 
The experimental evidence was obtained over a wide temperature range, from 
the critical temperature down to temperatures where the normal fluid 
fraction is only a few percent. 

At intermediate to high temperatures (relative to the critical temperature) the role of the mutual friction is
crucial, allowing energy exchange between normal fluid and superfluid.
The mutual friction depends on the relative velocity
of the two fluid components, the density of vortex lines, and  dimensionless
temperature-dependent friction coefficients $\alpha$ and $\alpha'$. 
Therefore, at large length scales, the turbulent normal fluid and superfluid tend 
to move together, locked by the mutual friction, a situation referred to
in the literature as ``co-flow" to distinguish it from ``counterflow" (see
Section~\ref{sec:discussion}). The situation in $^3$He-B is different
because the normal component is so viscous that its flow is 
laminar in all experiments; unlike in $^4$He, it plays no dynamical role
and simply provides a friction to the motion of the vortex lines. 
Nevertheless, even in $^3$He-B, a Kolmogorov turbulence is predicted
\cite{vinen-2005,lvov-nazarenko-volovik-2004}
when the cascading dynamics is significantly faster than the dissipative 
action arising from the mutual friction.
In both $^4$He and $^3$He-B,  
the proportion of normal fluid becomes less and $\alpha,\alpha' \to 0$ as the
temperature is lowered, leaving what is effectively a pure superfluid. 

Kolmogorov quantum turbulence at very low temperature
was created in both $^4$He and $^3$He-B
by injecting charged vortex rings  
\cite{Walmsley2007,Walmsley2008}, and by oscillating grids and forks 
\cite{Bradley2006,Davis}. In all these experiments the
turbulence can be considered incompressible (the average speed 
of vortex lines and moving boundaries being much smaller 
than the speed of sound), facilitating the comparison with classical HIT.  
In a turbulent superfluid, the rms vorticity is usually identified 
\cite{VinenNiemela} as $\approx \kappa L$.  The prediction \cite{Stalp} that
Kolmogorov quantum turbulence decays as $L \sim t^{-3/2}$ for large $t$
was experimentally verified\cite{Smith,Walmsley2007} in $^4$He 
using second sound or ion techniques, and
in $^3$He-B using Andreev scattering \cite{Bradley2006,Baggaley-Tsepelin}.
The corresponding decay of the total kinetic energy scales as $t^{-2}$ at
large $t$.

The numerical simulations which contributed to the evidence of
the Kolmogorov $k^{-5/3}$ scaling (in both statistically-steady and
decaying regimes) modelled the superfluid using the
Gross-Pitaevskii equation  \cite{Nore,Kobayashi} (GPE) or
the Vortex Filament Method  
\cite{Araki,Baggaley-Shukurov,Baggaley-Walmsley,Baggaley-Laurie,Galantucci-heli,
Galantucci-zero} (VFM). VFM simulations also 
reproduced the observed $L \sim t^{-3/2}$ temporal decay of the
turbulence. At non-zero temperatures, the coupled Kolmogorov dynamics 
of normal fluid and superfluid was demonstrated using models ranging from
the coarse-grained HVBK equations 
\cite{Roche-Barenghi-Leveque} 
to a modified Leith model \cite{Lvov-Leith} to modified shell models
\cite{GOY,SABRA}. In particular, the HVBK model agreed with the striking
experimental verification \cite{Salort45law} in $^4$He of the so-called 
4/5 law of classical turbulence, a statement about the third moments of velocity
increments which can be derived exactly 
from the Navier-Stokes equation for HIT in the inertial range.
Remarkably, the Kolmogorov picture of classical turbulence is also able 
to capture in quantum turbulence the behaviour of the 
scaling exponents of the low-order velocity circulation moments 
\cite{Muller2021};
higher order moments are described by a bifractal model, as in 
classical turbulence \cite{Iyer2019}.


A snapshot of Kolmogorov quantum turbulence
computed in a periodic domain is quite similar to vortex tangle
displayed in Figure~\ref{fig:1}, 
consisting of a disordered tangle of vortex lines whose average radius
of curvature is of the order of magnitude of the average inter-vortex
distance $\ell$. In the statistical steady-state, the
vortex lines continually collide and reconnect at the rate of
$\approx \kappa L^{5/2}$ reconnections per unit time per unit volume 
\cite{Sherwin}.

It is important to remark that the appearance of the classical 
Kolmogorov scalings in a quantum fluid is limited to the 
"classical length scales" \cite{Skrbek2021}
corresponding to wavenumbers
$k_D \ll k \ll k_{\ell}$ (where $k_{\ell}=2 \pi/\ell$). Only in this
range it is possible for vortex lines to locally polarise (even if only
partially), effectively creating classical eddies that can undergo
the process of vortex stretching (vortex stretching on individual
vortex lines is prevented by the quantisation of the circulation).
Usually, this polarisation is poorly visible in images such as
Figure~\ref{fig:1} which do not display the orientation of the vortex lines,
but it becomes apparent by computing a suitably-defined coarse-grained
vorticity field \cite{Baggaley-Laurie,Baggaley-bundles}.
The coarse-graining procedure
reveals that bundles of vortex lines can spontaneously come together,
parallel to each other in the mist of the random background
of the other vortex lines, creating regions
of relatively large velocity and energy.
The remaining vortex lines, although containing most of the 
vortex length \cite{RocheBarenghi}, contribute
less to the energy because 
their velocity fields tend to cancel out. 
A similar effect \cite{Farge}
takes place in classical HIT, where tubular
regions of large enstrophy and energy
are responsible for the $k^{-5/3}$ spectrum, and the rest of the flow
is incoherent.

On the contrary, in the ``quantum length scale" \cite{Skrbek2021} 
range of wavenumbers 
$k \gg k_{\ell}$, the dynamics is non-classical, because it 
strongly depends on the quantisation of the circulation,
which has no classical analogue.
In this range, the energy spectrum scales as $k^{-1}$, which is the spectrum
of an isolated straight vortex (at length scales shorter than $\ell$,
the dominant velocity field arises from the nearest vortex, which is
effectively straight at this scale).

The scenario which we have described (different dynamics of the turbulence
at the classical and at the quantum length scales) is confirmed 
by studies of the statistics of velocity components. 
Measurements \cite{Paoletti}
in $^4$He performed using tracer particles smaller than $\ell$
(as well as numerical simulations which focussed on the velocity
at given points in space \cite{White,GalantucciSciacca2014}) revealed velocity distributions
which scale as $v^{-3}$ at large $v$; this power-law behaviour
disagrees with the Gaussian statistics of classical HIT. More careful
numerical \cite{Baggaley-stats} and experimental 
\cite{LaMantia} studies verified that, if the measurement region
(in time or space) extends to distances larger than $\ell$, 
classical Gaussian statistics are recovered.

The disordered vortex tangle such as that shown in
Figure~\ref{fig:1} suggests that quantum turbulence has a rich topology.
A recent study \cite{Cooper} has shown that  this is indeed the case.
In the statistical steady-state,
the continual vortex reconnections knot and unknot vortex lines,
sustaining a spectrum of vortex knots (closed vortex loops with non-trivial
topology, the simplest of which is the trefoil). Surprisingly,
the vortex tangle always contains some 
knots of very high order,
as well as a non-zero degree of linkage between vortex lines. The precise
relation between the topology and the geometry/dynamics of the
turbulence is still a mystery, but this result hints at energy implications
(vortex reconnections, which create and destroy knots,
represent kinetic energy loss in the form of sound 
radiation). It has even been suggested \cite{Kauffman}
that the decay of superfluid turbulence may follow particular 
topological pathways.

Finally, as in classical turbulence, intermittency 
corrections are responsible for deviations from the self-similar 
Kolmogorov statistics \cite{Rusaouen2017,Varga2018,Muller2021}. 


\subsection{Vinen turbulence}
\label{sec:Vinen}

If quantum turbulence were always of the Kolmogorov type described in
Section~\ref{sec:Kolmogorov}, we would conclude that,
since the turbulence contains enough quanta of circulation,
the classical limit is indeed recovered 
in agreement with Bohr's correspondence principle. 
But a different type of quantum turbulence, first envisaged 
by Volovik \cite{Volovik}, was experimentally identified in $^4$He 
by Walmsley and Golov \cite{Walmsley2008} and in $^3$He-B 
by Bradley {\it et al}. \cite{Bradley2006}.  This type of turbulence, 
which decays as $L \sim t^{-1}$ at large times (corresponding to the
 total kinetic energy that decays as $t^{-1}$) 
is called ``ultra-quantum" or {\it Vinen quantum turbulence}. Numerically,
the $L \sim t^{-1}$ decay was seen in simulations 
\cite{Baggaley-Walmsley} of Walmsley and Golov's experiment 
and in simulations of turbulence driven by a uniform normal 
flow \cite{Sherwin} (both using the VFM); it was also seen
in simulations of the thermal quench of a Bose gas \cite{Stagg} using
the GPE. Besides reproducing the $L \sim t^{-1}$ decay, these
simulations revealed that the energy spectrum of Vinen quantum
turbulence is different from that of Kolmogorov quantum turbulence in two
important respects: it lacks the concentration of
energy at the large length scales near $k_D$ typical of classical turbulence,
peaking instead near $k \approx k_{\ell}$, and it
scales as $k^{-1}$ for large $k$. 
The energy spectrum of Vinen quantum turbulence is thus reminiscent 
of the energy spectrum of a random gas of vortex rings of radius $R$,
which peaks at $k \approx 1/R$ and scales as $k^{-1}$ for $k>1/R$ 
(if the rings' radii are not the same but span a distribution of values,
then the peak near $1/R$ broadens). The absence of the $k^{-5/3}$ scaling
suggests that in Vinen quantum turbulence the energy transfer from small wavenumbers 
$\sim k_D$ to wavenumbers $\sim k_{\ell}$ is weak or essentially absent.

The lack of polarisation of the vortex lines
in Vinen quantum turbulence becomes apparent by computing
the coarse-grained vorticity, which is very small (i.e. vortex lines
tend to be randomly oriented with respect to each other). For this reason,
the local mesoscale helicity, $h(\zeta)$ (where $\zeta$ is the
arc length), which measures the non-local 
vortex interaction \cite{Galantucci-heli}, is smaller 
than in Kolmogorov quantum turbulence. For the sake of
illustration, the vortex lines of Figure~\ref{fig:1}
are colour-coded according to $\vert h(\zeta)\vert$:
a low value (blue colour) at a point along a vortex line 
means that, at that
point, the line moves with velocity predominantly due to the 
local curvature; a high value (yellow colour) means that the line's velocity is 
predominantly due to other vortex lines. 
The weak vortex interaction in Vinen quantum turbulence
is reflected in the rapid decay with distance of the velocity correlation
function, as found by Stagg {\it et al.} \cite{Stagg}.

A model which accounts for the generation and decay of both
Kolmogorov and Vinen turbulence based on the fluxes of energy at the
classical and the quantum length scales was presented by Zmeev {\it et al.}
\cite{Zmeev}.
The key ingredient to create Vinen turbulence is either
the lack of forcing at the large length scales (in the steady case)
or an initial condition that lacks sufficient energy at the large length scales 
(in the unsteady case).  The paradigm is the initial condition 
which is used to model the thermal quench of a Bose gas and the
formation of a condensate \cite{Berloff-Svistunov,Stagg}:
the phase is spatially random and the occupation number is
uniform in $k$-space: without any anisotropy, in three dimensional
flows there is no mechanism to create large flow structures
\cite{Biferale,Baggaley-inverse}.


\subsection{Quantum turbulence at small length scales}\label{sec:small}

In quantum turbulence, the physics of the small length scales deserves
special attention.  In general, the friction with the normal fluid 
damps the energy of the vortex lines. Small-scale vortex
structures such as small vortex rings or loops shrink and vanish, passing
their energy to the normal fluid which turns it into phonons (heat) via viscous
forces. Other vortex structures that are affected by the friction
are Kelvin waves (helical oscillations of the centerline 
of an unperturbed vortex). Kelvin waves are created by vortex
interactions and reconnections \cite{Kivotides}, as seen experimentally \cite{Fonda}, or simply
thermally \cite{BDV}. 
When two vortex lines collide\cite{Galantucci-reconnections},
immediately after the reconnection both lines acquire the shape of a cusp.
As the two cusps relax, Kelvin wavepackets are radiated away along the vortex 
lines, and the friction reduces their amplitude as they propagate.  
The frequency of a Kelvin wave of wavenumber $k$
is $\omega \approx \beta k^2$ (i.e. shorter Kelvin waves move faster),
where $\beta=\kappa/(4 \pi) \ln{(k a_0)}$ and $a_0 \approx 10^{-10}~\rm$ is the
vortex core radius. The amplitude of the Kelvin wave decays 
exponentially with time as $\exp{(-\alpha \beta k^2 t)}$ where
$\alpha$ is a temperature-dependent friction parameter. 
As a consequence, the geometrical
appearance of the turbulent vortex tangle changes with temperature:
at high temperatures the vortex lines are very smooth, but as the
temperature is reduced and the friction coefficient $\alpha$ decreases,
the vortex lines display cusps, kinks and high frequency Kelvin waves
\cite{Tsubota-nofriction}.
 
Since the superfluid has zero viscosity, the classical definition of
Reynolds number does not apply. However, using the same argument for which
the Reynolds number measures the ratio of inertial to viscous forces
in the Navier-Stokes equation, a superfluid Reynolds number ${\rm Re}_s$ can be
defined by the ratio of inertial and friction forces.
One finds \cite{Finne} 
${\rm Re}_s=(1-\alpha')/\alpha$, where $\alpha$ and $\alpha'$ are
respectively the dissipative and non-dissipative friction coefficients.
Note that ${\rm Re}_s$ depends only on temperature, not on $U$ or $D$.
Since $\alpha$ and $\alpha'$ tend to zero for $T \to 0$, the limit
${\rm Re}_s \to \infty$ corresponds to the temperature $T \to 0$.

The lower the temperature, the more freely the 
Kelvin waves propagate, and the further the distribution of curvatures 
extends to large values. Recent numerical simulations \cite{Galantucci-zero} 
which carefully resolved numerically all Kelvin waves excited 
in the turbulence showed that, as ${\rm Re}_s$ increases,
the rate of kinetic energy dissipation arising from mutual friction, 
$\epsilon$, at first decreases, then 
it flattens and becomes constant, as for the classical dissipation
anomaly.  As in classical turbulence, the generation of small-scale vortex 
structures as the turbulence becomes more intense prevents the dissipation 
from vanishing, unlike what happens in laminar flows. 
This  is another remarkable similarity between classical and quantum 
turbulence.  Being a property of the small length scales, this property
applies to both Kolmogorov quantum turbulence and Vinen quantum turbulence
\cite{Galantucci-zero}. 

If the temperature is further reduced, a different route to
dissipate kinetic energy becomes possible, as discovered by
Svistunov \cite{Svistunov}.  The nonlinear
interaction of finite-amplitude Kelvin waves creates shorter and 
shorter waves which rotate more and more rapidly. 
This Kelvin wave energy cascade can transfer energy
to length scales much smaller than $\ell$, until they
are sufficiently small that radiation of sound (phonons) takes place.
Vinen \cite{Vinen-phonons}
estimated that in $^4$He the crossover from friction dissipation to
sound dissipation occurs at approximately $T \approx 0.5~\rm K$ for
$L \approx 10^{10}~\rm m^{-2}$ (the precise
value depending on whether it is dipole or quadrupole radiation).

At low temperatures, quantum turbulence may thus contain two
energy cascades: a Kolmogorov cascade
of bundled vortices (analog to classical eddies) in the range
$k_D \ll k \ll k_{\ell}$,
and a Kelvin wave cascade on individual vortex lines for $k \gg k_{\ell}$.
Large GPE simulations of quantum turbulence show that the two
cascade are separated by a bottleneck region \cite{diLeoni} and verified
\cite{Krstulovic-cascade} the predicted \cite{Lvov-Nazarenko-cascade} scaling 
behaviour of the Kelvin cascade. Recently, experimental evidence in liquid helium
of the Kelvin cascade has been announced \cite{Makinen}.


\subsection{Strong turbulence}
\label{sec:strong}

In three-dimensional atomic condensates, direct 
non-destructive visualization of vortex lines has been achieved
only for one or two vortex lines at the time \cite{Serafini,SerafiniGalantucci}.
The comparison between quantum turbulence in atomic Bose-Einstein
condensates and in superfluid helium is therefore hindered by 
the lack of experimental techniques to visualize
the turbulence, measure velocity fluctuations, and determine
the vortex line density.  In two-dimensional
condensates, instead, vortices can be visualized and counted, but the
physics of two-dimensional turbulence is very different and beyond the
scope of this paper.  

Another significant difference between atomic condensates and
superfluid helium is that, currently, the ratio between the largest and the
smallest length scales (the system's size $D$ and the vortex core
radius $a_0$, which is of the order of the healing length)
is typically $D/a_0 \approx 10^2$. This value must be compared
to $D/a_0\approx 10^{10}$  in the largest turbulent $^4$He facility 
(the SHREK facility \cite{Rousset2014})
and $D/a_0 \approx 10^5$ in a small $5~\rm mm$ sample of $^3$He.
As a consequence, the range of $k$-space available to determine any
the scaling law of turbulence in atomic condensates is quite limited. 
A final difference between atomic
condensates and superfluid helium, again from the point of view of turbulence,
is that most atomic condensates are experimentally
confined by harmonic trapping potentials which create a non-uniform
density profile.  The recent development of box-trap potentials
\cite{Gaunt} now allows condensates with uniform
density, opening the way to better comparison with classical HIT and 
quantum turbulence in superfluid helium. 

\begin{figure}[!h]
{\includegraphics[width=0.4\textwidth]{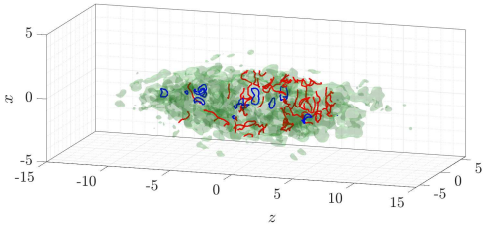}}
\caption{\label{fig:strong}
Strong turbulence in atomic Bose-Einstein condensate computed
using the GPE \cite{Holly} to simulate a shaken condensate. 
The red and blue lines mark the axes of
small vortex rings and U-vortices respectively. These vortex
structures coexist with the large density fluctuations (the green surface
is the isosurface of the density at an arbitrary value.}
\end{figure}


Although various techniques to create turbulent vortex lines in condensates
have been proposed and realized (phase imprinting
\cite{White,Cidrim}, rotation \cite{Kobayashi2007}, laser stirring
\cite{Kwon}, etc), until now the most successful strategy has been
shaking \cite{Henn2009} or oscillating 
the trap \cite{Navon2016}. Here we concentrate on results obtained 
by shaking the trap back and forth \cite{Henn2009}.

The most direct signature of turbulent dynamics in atomic
condensates is provided by two-dimensional density absorption 
images after expansion. These images have
revealed large density fluctuations and fragmentation,
and the lack of the expected inversion of
the aspect ratio of an initially cigar-shaped condensate \cite{Henn2009}.
But it is the momentum distribution $n(k)$ (determined, again,
by expanding the condensate) which has given the most important
physical information about the turbulence \cite{Bahrami,Garcia2022,Navon2016},
revealing inter-scale energy transfer 
\cite{Garcia2020,Navon2019} from small to large $k$.

Provided the temperature is sufficiently small compared to the critical 
temperature, the dynamics of
atomic condensates can be accurately simulated by the GPE. 
The computed \cite{Holly} momentum distribution scales as $n(k) \sim k^{-2.55}$ 
for large $k$, in good agreement with the experiments
($n(k) \sim k^{-2.60}$) in the range $k_D <k <k_{\xi}$ (where $k_D$ refers 
to the Thomas-Fermi radius and $k_{\xi}$ to the healing length).

Steeper momentum distributions ($n(k) \sim k^{-3.5}$, 
and, more recently \cite{Dogra}, $n(k) \sim k^{-3.2}$) 
were also observed in experiments in which the turbulence was excited 
by oscillating a box trap \cite{Navon2016}.
The large density fluctuations which are revealed by the numerical
simulations \cite{Holly} of the experiments have never been observed in
superfluid helium (where the typical Mach numbers are very small, 
as already remarked); the traditional reference problem - classical HIT -
is also incompressible. Fortunately, another classical reference problem 
is available: {\it weak turbulence} of waves. Zakharov's statistical theory
\cite{Zakharov} of interacting small-amplitude waves predicts
$n(k) \sim k^{-3}$, which is not far from values seen in condensates, 
particularly if one accounts for the finite-size of the system \cite{Dogra}. 
The difference between the scaling exponents observed in harmonic traps and in
box traps, and between experiments and wave turbulence theory,
may be due to the different density profile (which is uniform
in box traps), or the presence of vortex lines (which is not included
in Zakharov's theory).
In particular, the numerical simulations of shaken condensates \cite{Holly} 
show many small vortex rings and loops of
size comparable to the healing length which coexist with the large
density fluctuations, as shown in Figure~\ref{fig:strong}. These vortex
lines are not spread uniformly (they seem to concentrate
at the back of the moving condensate) and are oriented randomly, making
the turbulence non-homogeneous but isotropic. Issues worth exploring
are the applicability of classical weak turbulence theory to turbulent
systems (like these condensates) having very
large density waves, and the effect of the quantum pressure.

Compared to turbulent superfluid 
helium, the absence of long vortex lines in turbulent
condensates is striking (perhaps the reason 
is the way vortices nucleate from unstable solitons created
by the shaking near the boundary). Perhaps other techniques to excite
turbulence in atomic condensates may create vortex tangles more similar
to turbulent helium.
No scaling law is clearly visible in the
spectrum of the incompressible kinetic energy of a shaken condensate 
- the radius of curvature of the small vortex loops is too small compared
for the usual $k^{-1}$ range of isolated straight vortices to stand out;
the only recognizable scaling is the $k^{-3}$ behaviour near $k_{\xi}$
caused by the vortex core funnel. Nevertheless, the temporal 
decay of the vortex length scales as $t^{-1}$ for large $t$, which is
the signature of random vorticity typical of Vinen turbulence. Here the
decay is clearly due to sound radiation.

Considering together the properties which have been observed in turbulent
condensates and
comparing them to known turbulence types, we conclude that 
quantum turbulence in atomic condensates fits neither the
Kolmogorov type (for it lacks the characteristic polarised
vortex lines) nor the Vinen type (the random vorticity typical of
Vinen turbulence is dominated by strong density waves); it is also unlike 
the classical HIT scenario. Its closest classical analog is
clearly Zakharov's weak-wave turbulence. However it differs from it due to
the much stronger nonlinearity of the waves,  the additional 
presence of vortices whose effect on the waves' dynamics is still unexplored,
and the mechanism for energy loss (sound radiation). 
It seems that quantum turbulence in atomic condensates is clearly
a type of turbulence with its own characteristic properties. The
name {\it strong quantum turbulence} highlights the role played by
the large density fluctuations, making reference to the expression
used in the classical literature for turbulent
weakly-interacting waves.


\section{Discussion}
\label{sec:discussion}

The three types of quantum turbulence which we have reviewed
represent only a convenient way to organize experimental and numerical 
observations at this stage of progress. Their aim is to help 
identify and compare the underlying physics processes. 
As more quantum fluids and more quantum turbulent flows are discovered
and investigated, this classification may change, or more intermediate types
will become known.

The flexibility and versatility of the atomic condensates allow for the laboratory creation of more complex quantum fluids which may in turn support novel types of quantum turbulence.  One example are the dipolar atomic condensates,  in which the particles possess sizeable magnetic dipole moments\cite{Lahaye2009}.   The quantum fluid then gains a magnetic character and a response to imposed magnetic fields analogous to a classical ferrofluid,  leading to regimes of polarized and stratified quantum turbulence and a new modality to externally drive and control the fluid\cite{Bland2018}.  A second example are the multi-component condensates,  in which distinct condensates co-exist and interact such that the system embodies the quantum analog of a multifluid.   The fluids can either be miscible or immiscible depending on the atomic parameters, and the individual components can be independently addressed,  e.g. to controllably induce co-flow or counter-flow.   For both the dipolar and multi-component systems,  the additional interactions and degrees of freedom may provide additional channels for energy transfer and dissipation, as well as supporting distinct instabilities which may lead to the onset of turbulence.   Although quantum turbulence has yet to be experimentally probed in either system,  advances have enabled vortex states to be experimentally created in both \cite{Klaus2022,Schweikhard2004} such that extending to turbulent regimes is within reach.

The recognition of different
types of quantum turbulence may be particularly useful for problems 
which cannot be directly accessed in the laboratory, and numerical
modelling is constrained by scarce observations. 
Two examples of current astrophysical interest are worth mentioning. 
The first is neutron stars (pulsars). There is some evidence that the 
observed rotational glitches of these stars are related to quantum turbulence 
in the interior \cite{Haskell}.  Unfortunately
the only experimental information about rotating quantum
turbulence is from few studies of rotating thermal counterflow 
\cite{Swanson,PerettiGibert} in helium, and
thermal counterflow turbulence is an old problem still under
investigation (as we shall see at the end of this section). 
Numerically, rotating quantum turbulence is still at early stage
of investigation, and appears to be very different from classical
rotating turbulence \cite{Estrada}.
A better understanding of the 
spin-down and spin-up dynamics of quantum turbulence would help setting
up more efficient observational protocols for pulsar glitches.

The second example is dark matter. If it consists of light bosons, a possible
model would be a self-gravitating condensate described by the 
Gross-Pitaevskii-Poisson equation. Numerical simulations of dark matter
galactic haloes have revealed tangles of vortex lines with a Vinen-like
kinetic energy spectrum \cite{Mocz,Liu}, alongside huge density 
fluctuations which are even larger than in the strong turbulence of
atomic condensates.
Such model may provide information about micro-lensing which would help 
confirm or discard this approach to the dark matter problem.

Finally, we remark that the very first quantum turbulent flow which
was studied in Vinen's pioneering work - thermal counterflow 
in superfluid helium - is surprisingly hard to classify. 
This flow is created by an electric heater which deposits
a given heat flux into a sample of helium, resulting in the opposite
motion (counterflow) of the normal fluid component (away from the heater) 
and the superfluid component (towards it). If this counterflow velocity
exceeds a small critical value, a turbulent tangle of vortex lines 
is generated which limits the otherwise ideal heat conducting property 
of liquid helium. Thermal counterflow has therefore
important applications of cryogenic engineering. Although there is 
experimental and numerical
evidence that at small heat flux the turbulence is Vinen-like 
(decaying as $L \sim t^{-1}$), at large heat flux the decay is 
Kolmogorov-like\cite{Skrbek2003} (decaying as $L \sim t^{-3/2}$).
Because of the opposite motion of normal fluid and superfluid,
this quantum turbulent flow which has no classical analogy does
not fit our classification, indeed it is still the subject of 
investigations \cite{GuoKanai}. 

In conclusion, the legacy of Vinen's early experiments on the quantisation
of circulation \cite{Vinen-quantum} and heat conduction
in liquid helium \cite{Vinen-counterflow} has been profound. 
Taken together, these two experiments
have opened new doors to out-of-equilibrium quantum systems and
the meaning of the classical limit which are currently expanding in directions
which he could not have anticipated.

\begin{acknowledgments}
C.F.B. is indebted to W.F. Vinen for support and discussions
over many years. The financial support of UKRI 
(UK Research and Innovation) under
grant ``Quantum simulators for fundamental physics" (ST/t006900/1) is 
acknowledged.
\end{acknowledgments}





\end{document}